\documentclass[preprint,showpacs,aps,amsmath,amssymb]{revtex4}
\usepackage{times}
\usepackage{mathptmx}
\usepackage{graphicx}
\usepackage{bm}

\begin{document}

\title{Mechanisms in Adaptive Feedback Control:\\
Photoisomerization in a Liquid}

\author{Kunihito Hoki}
\author{Paul Brumer}
\affiliation{Chemical Physics Theory Group, Department of Chemistry,
 and Center for Quantum Information and Quantum  Control, 
  University of Toronto, Toronto, Canada M5S 3H6}

\date{\today}

\begin{abstract}
The underlying mechanism for Adaptive Feedback Control in the 
experimental photoisomerization of NK88 in methanol is exposed 
theoretically. With given laboratory limitations on laser output,
the complicated electric fields are shown to achieve their
targets in qualitatively simple ways. Further, control over the 
{\itshape cis} population 
without laser limitations reveals an incoherent pump-dump
scenario as the optimal isomerization strategy. In
neither case are there substantial contributions from quantum 
multiple-path interference or from nuclear wavepacket coherence.
Environmentally induced decoherence 
is shown to justify the use of a simplified theoretical model.
\end{abstract}


\maketitle

Experimental results on the quantum control of molecular 
processes~\cite{bsbook,dantus_2004} fall into two categories: those
designed to explore the utility of a particular coherent control
scenario, and those that use adaptive feedback~\cite{judson_prl_1992,
bardeen_cpl_1997, levis_science_2001, brixner_cpc_2003, vogt_2004,
bucksbaum_2004} to attempt automated optimal control of a target
process.  The former have well understood mechanisms but are
scenario-specific, whereas the latter are generally applicable but
have thus far provided very limited insight into the mechanisms by
which control is achieved.  Indeed, it is widely recognized that
extracting the mechanism of control from an optimal control experiment
is the central challenge in this research
area~\cite{daniel_science_2003}.  In this letter we (a) expose the
simple mechanism underlying control in a recent~\cite{vogt_2004} 
{\itshape  trans--cis} isomerization control experiment, and (b) demonstrate 
that removing 
experimental restrictions on laser frequency and amplitude 
exposes modified versions of well known coherent control
scenarios as the dominant control mechanism.  

Gerber's seminal experiment~\cite{vogt_2004} shows successful
optimization of the {\itshape trans} to {\itshape cis} isomerization
of 3,3'-diethyl-2,2'-thiacyanine iodide (NK88)
as well as suppression of the {\itshape trans} to
{\itshape cis} transition, depending upon laser pulse shapes.  As is
typical of these experiments, the optimal pulse shapes in both cases
are complex functions of phase, frequency and time, and differ
substantially for the two targets.  Below we obtain quantitative
agreement with these results.
Further, we do so, despite the complexity of the system,
using a one degree of freedom model coupled to a bath.
Indeed, it is precisely the presence of environmentally induced
decoherence that simplifies the control problem.

Consider a model consisting of the system with Hamiltonian
$H_{\text{S}}$, the bath $H_{\text{B}}$, system--bath coupling
$H_{\text{SB}}$, and system--electric field coupling described in the
dipole approximation. The total Hamiltonian is given by
\begin{equation}
  H=H_{\text{S}}+H_{\text{B}}+H_{\text{SB}}-\mu E\!\left(t\right),
\end{equation}
where $H_{\text{S}}$ describes the isomerization process via a one
dimensional reaction coordinate $\phi$,  $H_{\text{B}}$ represents all
other degrees of freedom, $\mu$ is the transition dipole moment, and
$E\!\left(t\right)$ is the incident electric field at time $t$.

In terms of the two participating 
electronic states, the system Hamiltonian is given by:
\begin{equation}
  H_{\text{S}}=
  \begin{pmatrix}
    K + V_{\text{g}}\!\left(\phi\right) &
    V_{\text{ge}}\!\left(\phi\right) \\
    V_{\text{eg}}\!\left(\phi\right) &
    K + V_{\text{e}}\!\left(\phi\right)
  \end{pmatrix},
\end{equation}
where  $K=-\frac{\hbar^2}{2m}\frac{\partial^2}{\partial\phi^2}$ is the
kinetic energy, $V_{\text{g}}\!\left(\phi\right)$ and
$V_{\text{e}}\!\left(\phi\right)$ are the 
ground and excited electronic state potential surfaces, and
$V_{\text{ge}}\!\left(\phi\right)=V_{\text{eg}}\!\left(\phi\right)$ is
the coupling potential between ground and excited states.  In the adiabatic
representation the ground state potential is a double well~\cite{vogt_2004}.  
Populations reported below are of the eigenstates of the full molecular
Hamiltonian $H_{\text{S}}$.
The simplest dynamics takes place by photoexcitation from the {\itshape trans}
configuration to the excited electronic state followed by
de-excitation to the {\itshape cis} and {\itshape trans} ground state
via system--bath coupling.

In the case of NK88, the effective mass is $m=5$\,amu\,\AA$^2$,
$V_{\text{g}}\!\left(\phi\right)=A_0\left(1-\cos\phi\right)$,
$V_{\text{e}}\!\left(\phi\right)=A_1+A_2\cos\phi$,
$V_{\text{eg}}\!\left(\phi\right)=1000$\,cm$^{-1}$, where
$A_0$, $A_1$ and $A_2$ are 15900\,cm$^{-1}$, 17500\,cm$^{-1}$ and
7500\,cm$^{-1}$, respectively.  These, and other parameters below were
obtained by a fit to experimental results.
The eigenvalues $\lambda_i$ and
eigenvectors of the system are calculated by diagonalizing the
molecular Hamiltonian matrix represented on a grid, where periodic
boundary conditions at $2\pi$ are imposed on the system.  The
transition dipole moment $\mu$ is expressed in terms of the two
electronic states as a 2 $\times$ 2 dimensional matrix with zeroes on
the diagonal, and off-diagonal elements
$\mu_{\text{ge}}\!\left(\phi\right)=10$\,Debye.  This corresponds to
the oscillator strength $f\simeq 1$ in the Franck--Condon region of
the {\itshape trans} configuration.

The bath is described as a set of  harmonic oscillators of frequency 
$\omega_{\alpha}$ and the system--bath coupling is
$H_{\text{SB}}=Q\sum_{\alpha}\hbar\kappa_{\alpha}
\left(b_{\alpha}^{\dagger}+b_{\alpha}\right)$,
where $b_{\alpha}^{\dagger}$ and $b_{\alpha}$ are the creation and
annihilation operators pertaining to the $\alpha$th harmonic
oscillator.  The operator $Q$ is a diagonal $2\times 2$ matrix with
$\cos\phi$ on the diagonal, and the coupling constant
$\kappa_{\alpha}$ and spectrum of the bath are chosen in accord with 
an Ohmic spectral density
$ J\!\left(\omega\right)
  = 2\pi\sum_{\alpha}\kappa_{\alpha}^2
  \delta\!\left(\omega-\omega_{\alpha}\right)
  = \eta\omega e^{- \omega / \omega_c}$,
where the strength of the system--bath coupling is determined
by the dimensionless parameter $\eta=5$, and
$\omega_c=450$\,cm$^{-1}$.  Given these parameters, the electronic
dephasing time around the Franck--Condon region of the
{\itshape trans} configuration is $\sim 10$\,fs, and virtually complete
relaxation from
excited {\itshape trans} to stable {\itshape trans} and {\itshape cis}
occurs within 5\,ps. The former is a typical characteristic dephasing
time whereas the latter is chosen to agree with experiment.

The dissipative dynamics of the system was evaluated using the 
Redfield equation with secular 
approximation~\cite{blum_plenum_1981}. The
evolution of diagonal elements $\rho_{ii}\left(t\right)$
of the system density matrix is given by
\begin{align}
  \label{eq:propa_diag}
 \frac{\partial}{\partial t}\rho_{ii}\left(t\right)
 &= - i\frac{E\!\left(t\right)}{\hbar}
     \sum_m\left[\rho_{im}\!\left(t\right)\mu_{mi}
       -\mu_{im}\rho_{mi}\!\left(t\right)\right] \notag\\
 &+\sum_{j \neq i} w_{ij} \rho_{jj}\left(t\right) - 
\rho_{ii}\left(t\right)\sum_{j \neq i}w_{ji},
\end{align}
where the transition probability is
$w_{ji}=\Gamma_{ijji}^{+}+\Gamma_{ijji}^{-}$ and where each index denotes 
a state of the system, including the electronic and vibrational
quantum numbers.  Here,
$\Gamma_{ljik}^{-}=\left(\Gamma_{kijl}^{+}\right)^*$ and
\begin{align}
  \label{eq:gamma}
  &\Gamma_{kijl}^{+}
  = \frac{1}{2\pi}Q_{lj}Q_{ik}
  \int_{0}^{\infty}d\tau \int_{0}^{\infty}d\omega
  J\!\left(\omega\right) \cdot \notag \\
  & \quad\cdot
  \left\{
    \left[\bar{n}\!\left(\omega\right)+1\right]
    e^{-i\left(\omega_{ik}+\omega\right)\tau}
    + \bar{n}\!\left(\omega\right)
    e^{-i\left(\omega_{ik}-\omega\right)\tau}
  \right\},
\end{align}
where $\bar{n}\!\left(\omega\right)
=\left\{\exp\!\left(\hbar\omega/k_bT\right)-1\right\}^{-1}$ is the
Bose distribution, $T$ is a temperature, and
$\omega_{ji}=\left(\lambda_j-\lambda_i\right)/\hbar$.  After some
algebra, we obtain
\begin{equation}
 w_{ji} = \begin{cases}
	   \left|Q_{ji}\right|^2 J\!\left(-\omega_{ji}\right)
	   \left[\bar{n}\!\left(-\omega_{ji}\right)+1\right]
	   &\mspace{13mu}\text{for } \omega_{ji}<0 \\
	   \left|Q_{ji}\right|^2
	   J\!\left(\omega_{ji}\right)\bar{n}\!\left(\omega_{ji}\right)
	   &\mspace{13mu}\text{for } \omega_{ji}>0
	  \end{cases}.
\end{equation}
The evolution of the off-diagonal elements is described as
\begin{align}
  \label{eq:propa_offdiag}
  \frac{\partial}{\partial t}\rho_{ij}\left(t\right)
  &= -i\omega_{ij}\rho_{ij}\!\left(t\right)
  -\gamma_{ij}\rho_{ij}\!\left(t\right)  \notag\\
  &- i\frac{E\!\left(t\right)}{\hbar}
  \sum_m\left[\rho_{im}\!\left(t\right)\mu_{mj}
    -\mu_{im}\rho_{mj}\!\left(t\right)\right],
\end{align}
with dephasing rate $\gamma_{ij}$:
\begin{equation}
 \gamma_{ij}
  = \sum_{k}\left(\Gamma^+_{ikki}+\Gamma^-_{jkkj}\right)
  - \Gamma^+_{jjii} - \Gamma^-_{jjii}.
\end{equation}
The resultant vibrational dephasing time within the excited electronic
state is $\approx$ 15 fs.

To model the adaptive feedback experiment, the electric field comprises
128 frequency values, where the phases of each frequency component
are the optimization parameters; the frequency width is
200\,cm$^{-1}$, and the time width is 2\,ps. Specifically,
the electric field function is therefore taken to be
\begin{align}
  E\!\left(t\right)
  &=\sum_{i=0}^{127} A \exp\left[
    -\left(\frac{t-t_0}{2\Delta t}\right)^2
    -\left(\frac{\Omega_i-\Omega_0}{2\Delta \Omega}\right)^2
    \right] \cdot \notag\\
  & \cdot \cos\left(\Omega_i t + \Theta_i \right),
\end{align}
where $A$=5 MV/m, $t_0=\Delta t = 2$\,ps,
$\Omega_0=25000$\,cm$^{-1}$, $\Delta\Omega=200$\,cm$^{-1}$,
$\Omega_i=24800+3.125\times i$ in cm$^{-1}$, and $\Theta_i$ are the
optimization parameters.  Note that in accord with
experiment~\cite{vogt_2004} the field is optimized by varying the
phases $\Theta_i$ using an evolutionary algorithm, and the field
amplitude, as well as the overall frequency width of the pulse, are
constrained.  Further, the algorithm is designed to simulate
experimental conditions~\cite{baumert_apb_1997}, where the population
size is 60, 10 survivors are selected from the generation, each of
which has 4 children by mutation and 1 child by crossover, and
yielding again 60 individuals for the next generation.

In our second study, the experimental frequency and amplitude
constraints were lifted and optimal field was evaluated by using a 
monotonically
convergent algorithm~\cite{ohtsuki_jcp_1999}.  In both cases the 
optimization was carried out with a standard penalty on the laser
power, and the initial condition $\rho_{i,j}\!\left(0\right)$ was set to
thermal equilibrium at temperature $T=300$\,K.  At this temperature
the initial {\itshape cis} population is negligible.

Below, the time-dependent population of the stable {\itshape trans}, is
defined as the projection of $\rho\!\left(t\right)$ onto the lowest 49
states localized around $\phi =0$ and the {\itshape cis} population as
the lowest 23 states localized around $\phi=\pi$.  The excited state
population is the remainder.

\begin{figure}[htbp]
  \includegraphics{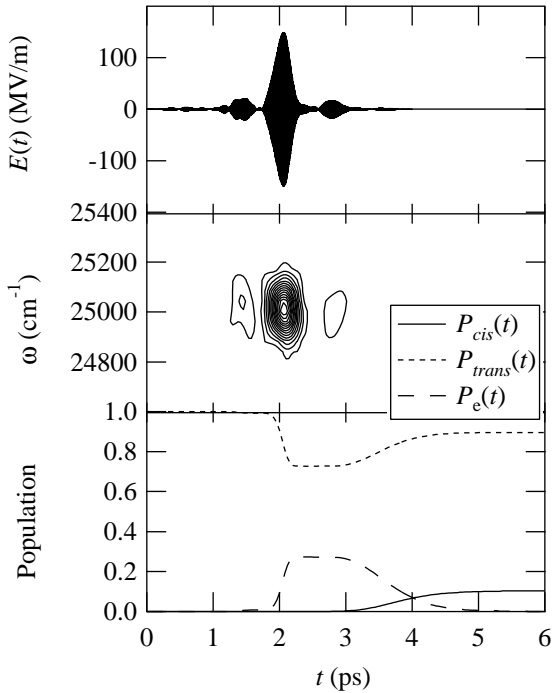}
  \caption{\label{fig:ga_max}
    Time evolution under the optimized electric field with restriction
    on frequency and amplitude.  Top panel: an electric field obtained
    by 64 iterations from random-initial phases $\Theta_i$.
    Middle panel: time--frequency resolved spectrum.  Bottom panel:
    time evolution of populations. }
\end{figure}

Figure~\ref{fig:ga_max} shows the isomerization dynamics under the
optimized electric field restricted in frequency and amplitude.  In
the experiment~\cite{vogt_2004}, the target was chosen as the
ratio of the transmission change $\Delta T$ at 400\,nm to that at 460\,nm, 
measured at 20\,ps, assumed to be proportional to 
number of {\itshape cis} molecules created and number of {\itshape trans}
molecules excited, respectively~\cite{vogtprivate}.  The 
target is modeled computationally as  
the ratio of population of stable {\itshape cis} to the laser pulse area,
$\mu_{\text{ge}}\int \left|E\left(t\right)\right| dt / \hbar$, at a
target time 20\,ps.  Here the pulse area would be approximately proportional 
to the number of {\itshape trans} molecules excited, if the 
nuclei were fixed and two electronic states participated in the process.
Its use corrects somewhat for the presence of decoherence in this one
dimensional model, relative to the multi-dimensional 
decoherence-free excitation experimentally.
The electric field is optimized by 64 iterations from randomly
initialized phases $\Theta_i$.  The resultant electric field 
(Fig.~\ref{fig:ga_max}), in accord with experiment, is seen
to have a peak at $\sim 1.9$\,ps, and is considerably sharper than the 
laser 2\,ps envelope for each frequency component, implying that the
optimization has yielded a nearly equal set of phases $\Theta_i$.
Figure~\ref{fig:ga_max} also shows small peaks around the main
peak, giving an overall structure  that is almost quantitatively the same as
the experimental result~\cite{vogt_2004}.  The bottom panel of
Fig.~\ref{fig:ga_max} shows the isomerization dynamics; the system
is seen to be excited by the electric field at $t\sim 2$\,ps, with
maximum $P_{\text{e}}\!\left(t\right)\simeq 0.4$.  After 
photoexcitation, the system relaxes into stable {\itshape trans} and
{\itshape cis} due to the system--bath coupling, giving
a final {\itshape cis} probability of 0.17.  The optimized electric
field gives a target value of 0.53, while an unoptimized electric
field gives a 0.7 times smaller value of 0.36, in agreement with experiment.

Further studies shows that the small peaks in $E(t)$ are of little 
relevance.  That is, after 200 iterations these small peaks almost
disappear.  Thus, the mechanism underlying the experimentally observed
control is
{\itshape efficient photo-excitation} under dissipative conditions,
balancing the time scale for excitation and wave packet motion with
the ongoing decoherence.

\begin{figure}[htbp]
  \includegraphics{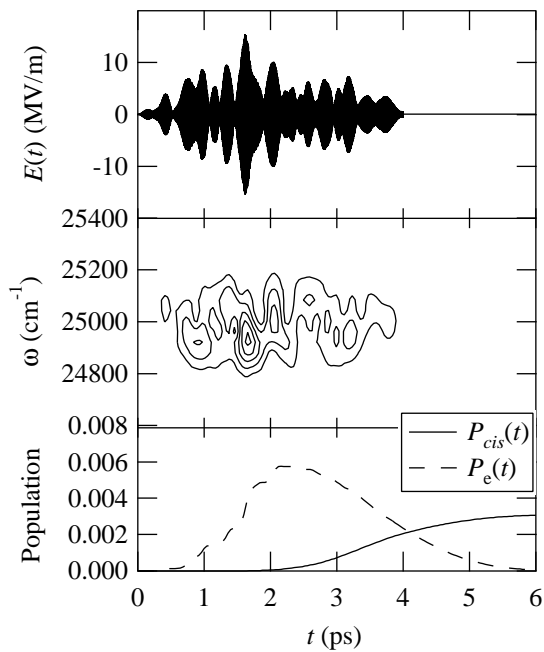}
  \caption{\label{fig:ga_min}
    Time evolution of the system under the optimized electric field
    that is restricted in frequency and amplitude.  In this case, the
    target of the control is minimization of stable {\itshape cis}.
    Top panel: optimized electric field.  Middle
    panel: time--frequency resolved spectrum.  Bottom panel: time
    evolution of populations. }
\end{figure}

Experimental results were also presented~\cite{vogt_2004} for the
case of ``no {\itshape cis} population'', again using fields
constrained in amplitude and frequency and using the transmission ratio as the
target. The analogous computational
result is shown in  Fig.~\ref{fig:ga_min}, in excellent agreement with
experiment. Specifically, the pulse is now delocalized in time, rather
complicated in form, and of duration longer than the dephasing time
between the two electronic states.  The maximum amplitude is
considerably lower than that shown in Fig.~\ref{fig:ga_max}.
Examination of the populations (bottom panel, Fig.~\ref{fig:ga_min}),
shows that $P_{\text{e}}\!\left(t\right)$ reaches only 0.006, far
lower than that in Fig.~\ref{fig:ga_max}.  Thus, despite the
complexity of the pulse, its entire purpose is to ensure that there is
no excitation of the initial {\itshape trans} species! 
However, the optimized
electric field in our case gives a 0.05 target value,  $\approx 4$ times 
smaller than the experiment, resulting from the fact that our model 
does not take into account competing 
processes such as excitations to other electronic states and the
large number of nuclear vibrational modes.

The robustness of our results were checked by performing
various alternate computations. 
For example, using the target "ratio of
created {\itshape cis} to depleted
{\itshape trans}" was not useful since this quantity is always unity at 
20\.ps. This is because in this model
the molecule is, due to relaxation by 20\.ps, either
stable {\itshape trans} or stable {\itshape cis}. Choosing a 
slower relaxation rate, so that excited population remains at
20\,ps, also did not improve the results. Here the target maximization showed 
an electric field moved to as early a time as possible 
to wait for the slow relaxation, or as late as possible
in the case of minimization to avoid relaxation to {\itshape trans} and
{\itshape cis} molecule.  Neither agree with experiment. 
However, using of a simpler
target, that is population of stable {\itshape cis}, gives
qualitatively the same fields as shown in Figs~\ref{fig:ga_max}
and~\ref{fig:ga_min}, but dissimilar ratios of target improvement relative 
to the unmodulated pulse. Hence we are confident that the essential physics is
contained in the simulation presented above.

The experimental results optimize the target within the 
restricted frequency range and intensity described above.  
Ideally, however, adaptive feedback control desires the
{\itshape optimal} result, which would require unrestricted laser
equipment.  To examine one such optimal solution we repeated
the adaptive feedback studies with penalties on the power, but with
no frequency restriction on the laser and with the
population of stable {\itshape cis} used directly as the target.

\begin{figure}[htbp]
  \includegraphics{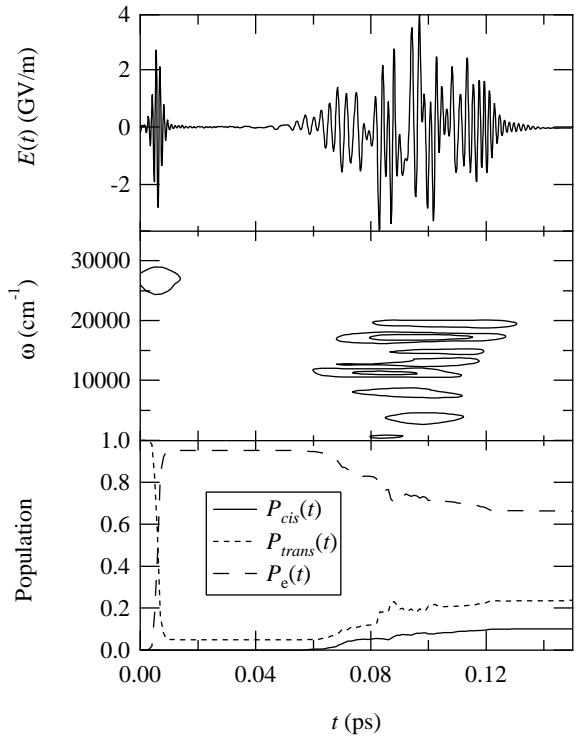}
  \caption{\label{fig:opt}
    Short-time evolution of the system under the fully optimized pulse.
    After 0.15\,ps, the electric field is essentially zero.
     Middle panel:
    time--frequency resolved spectrum of the pulse.  Bottom panel:
    time evolution of populations. }
\end{figure}


In the second of these target cases, the suppression of
{\itshape trans} to {\itshape cis} isomerization, the optimal solution
obtained was $E(t) = 0 $, i.e. no excitation. By contrast, the fully
optimal pulse for producing {\itshape cis} from {\itshape trans} is
shown in Fig.~\ref{fig:opt} where,  for computational convenience, the
target time is 5\,ps.  The optimal electric field seen to consist of a
pump pulse (from 0 to 0.01\,ps), and dump pulses (from 0.05 to
0.13\,ps). The dump pulse has several frequency components, and is
resonant with the deexcitation between electronic states around
$\phi=2.1$\,rad. After the pump pulse, $P_{\text{e}}\!\left(t\right)$
reaches almost 1, and after the dump pulse, $\sim 0.1$ of population
is transfered from $P_{\text{e}}\!\left(t\right)$ to
$P_{\text{\itshape cis}}\!\left(t\right)$.  Subsequently (not shown), the
remaining excited component relaxes 
due to the system--bath coupling giving a final {\itshape cis}
probability of 0.36, far higher than the {\itshape} cis 
probability obtained when this computation was repeated with the laser frequency
restrictions above. 

Note that unlike the paradigmatic
pump--dump coherent control scenario~\cite{tannorrice,bsbook}, or the
previously proposed eigenstate based {\itshape cis}--{\itshape trans}
isomerization mechanism~\cite{danny} the pump and dump steps are
{\itshape not} coherently related since the system is decohered
between pulses.  Hence the optimal mechanism
in this case is desirable excitation of an excited state population,
delocalizing and decohering of the excited state wavepacket, and 
desirable de-excitation of the decohered excited state population.
The delay between the pump and dump steps results from the spreading
of density into the {\itshape cis} region.

It should be noted that the optimal pulse has almost zero amplitude
after 0.15\,ps, since it is very difficult to try to make more
{\itshape cis} after the dump process.  That is, a second
photo-excitation step from the stable {\itshape trans} to the excited
state is not useful, since the excited component can not relax to stable
{\itshape cis} within the few alloted ps.  Also, due to the dephasing
of the system, the excited component spreads rapidly into whole $2\pi$
range of $\phi$.  Any attempt at a second photo-deexcitation from these
excited component to stable {\itshape cis} would also be ineffective
since it would be accompanied by compensatory excitation from the populated
{\itshape cis} to the excited state. 

In summary, this work has (a) successfully exposed the simple
underlying mechanisms associated with the complex experimental 
results of an adaptive, condensed phase,
feedback experiment, (b) demonstrated the
role of frequency limitations in the experimental laser
wavelengths and the concomitant emergence of an incoherent pump--dump
scenario when these restrictions are lifted, with {\itshape cis} 
population as the target, and (c) demonstrated the
utility of the simplest of models, one dimensional motion plus decoherence.
The utility of such models derives from the fact that the decoherence is
fast and that the measurement of isomer identity implicitly ignores all
degrees of freedom but one (the angle $\phi$). Hence, one dimensional
motion plus decoherence is formally the proper description. 
Note, however, that a appropriate representation of
the decoherence is necessary to achieve the quality of results shown
here.  Since most liquid phase control experiments will be of a
similar nature, simple 
control models of this type may  be justified generally by the role and
presence of decoherence.  However, if decoherence is slower than the
characteristic molecular dynamics then the
problem is more complex and dynamics in many degrees of freedom 
must be explicitly controlled.

Acknowledgment: We thank Professor Gustav Gerber and Drs. Gerhard Vogt and
Gerhard Krampert for discussions on their experiment.
This work has been supported by Photonics Research Ontario and the
Natural Sciences and Engineering Research Council of Canada.

\bibliography{cistransoptimal3}

\end{document}